\documentclass[epj]{svjour}

\usepackage{graphicx}

\begin{document}

\title{Internet's Critical Path Horizon}

\author{Sergi Valverde \inst{1} \and Ricard V. Sol\'{e} \inst{1,2}}

\institute{ICREA-Complex Systems Lab, Universitat Pompeu Fabra, 
Dr Aiguader 80, 08003 Barcelona, Spain  
\and 
Santa Fe Institute, 1399 Hyde Park Road, New Mexico 87501, USA}

\date{\today}
\abstract{Internet is known to display a highly heterogeneous 
structure and complex fluctuations in its traffic dynamics. Congestion 
seems to be an inevitable result of user's behavior coupled to the 
network dynamics and it effects should be minimized by choosing 
appropriate routing strategies. But what are the requirements of routing depth in order to optimize 
the traffic flow? In this paper we analyse the behavior of Internet traffic 
with a topologically realistic spatial structure as described in a previous study 
(S-H. Yook et al. ,Proc. Natl Acad. Sci. USA, {\bf 99} (2002) 13382). The model 
involves self-regulation of packet generation and different levels of routing depth. It is shown that 
it reproduces the relevant key, statistical features of Internet's traffic. Moreover, 
we also report the existence of a critical path horizon defining a transition from 
low-efficient traffic to highly efficient flow. This transition is actually a direct consequence of the 
web's small world architecture exploited by the routing algorithm. Once routing tables reach the network diameter, 
the traffic experiences a sudden transition from a low-efficient to a highly-efficient behavior. It is conjectured that 
routing policies might have spontaneously reached such a compromise in a distributed manner. Internet would thus 
be operating close to such critical path horizon.
}

\maketitle

\mail{svalverde@imim.es\\ricard.sole@upf.edu}

\section{Introduction}

The efficient performance of any communication network is jeopardized by 
congestion problems, which often show up in unpredictable ways. This seems 
the case, for example, of so called Internet storms \cite{Huberman}. Such problems 
were early identified in different types of engineered networks. Norbert Wiener for 
example mentions that ``a switching service involving many stages and designed 
for a certain level of failure shows no obvious signs of failure until the 
traffic  comes up to the edge of the critical point, when 
it goes completely into pieces, and we have a catastrophic traffic jam'' \cite{Wiener}. 
These observations allow to formulate a number of key questions concerning 
communication nets, such as: How do critical traffic levels are reached? What is the 
nature of these thresholds? How appropriate routing algorithms modify this behavior? 
Are there optimal routing strategies? 

\begin{figure}
\begin{center}
\includegraphics[scale=0.45]{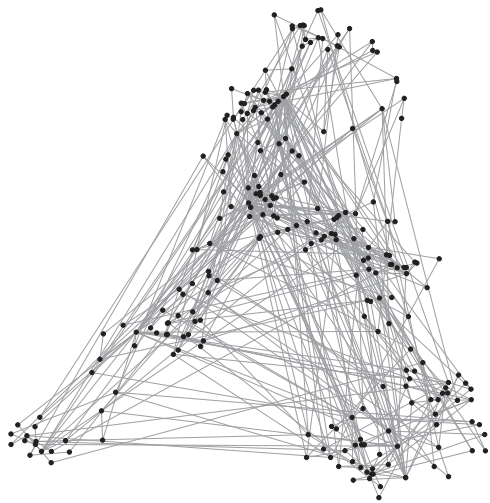}
\vspace{0.2 cm}
\caption{Small Internet-like network (see text for generation method description). 
The parameters have been set to the position corresponding 
to Internet in the $(D_f, \alpha, \sigma)$- phase space: M= 250, 
 $<k>=4$, $D_f=1.5$, $\alpha=1$, $\sigma=1$.
} 
\label{fig2}
\end{center}
\end{figure}

Modeling Internet dynamics has been an active area over the last decade. The approaches 
include detailed simulations \cite{Ogielski}, simple statistical models and verbal models 
\cite{WillingerPNAS}. In this context, in \cite{SoleValverde2001} we investigated the existence of a 
jamming phase transition between the free phase and the 
congested phase in a model of network traffic over regular 
meshes. This phase transition depends on traffic density. It was 
conjectured that large-scale network traffic self-organizes near 
the critical point of the transition, which is linked to high 
network efficiency and unpredictability \cite{ValverdeSole2002}. 
At the critical regime, the distribution of congestion duration lengths scales 
as a power-law \cite{Fukuda}\cite{ValverdeSole2002}.

Such a self-organized scenario is reinforced by a recent study 
suggesting that Internet fluctuations are a consequence of the 
internal dynamics of the system \cite{Fluctuations}. Against previous claims,  
there is mounting empirical support that network traffic heterogeneity is a consequence 
of collective dynamics and not because of the high variability injected by external 
sources.

In this paper we will show that Internet is able to route efficiently its inner flow of packets because a special 
combination of local routing rules and a particular network architecture.
As we will see, Internet routing reaches an optimal, low-cost 
traffic flow as a result of a trade-off between random and 
deterministic routing schemes.

\section{Modelling Internet's traffic}

Following previous 
 approaches \cite{SoleValverde2001}\cite{ValverdeSole2002} let us consider 
 network defined on a graph $\Omega$ with M nodes (figure 1). The number of 
 links connecting a given node with another (irrespective of their characteristics) 
 will be indicated as $k_i$. The subset of closest nodes of $s_i  \in \Omega$
 is $C_i$. Only a fraction of fixed $\rho < (0,1]$ nodes is 
 selected as source and destination traffic endpoints (hosts). Host 
 locations are chosen randomly. The other $(1-\rho )M$ nodes can only store 
 and forward incoming messages (routers).
 
Both host and routers can route only one packet at a time (the 
routing policy is described below). Each node $s_i$ is 
provided with a finite queue of packets waiting for communication 
resources (free links). Queues are not allowed to store more than $H$ packets 
simultaneously. New packets arriving at an already saturated queue 
will be simply removed from the system. This provides a very simple 
dissipative mechanism useful for recovering from heavy traffic 
congestion. If $n(s_i,t)$ is the number of packets at $s_i$, the total 
number of packets in the system will be 

\begin{equation}
N(t) = \sum_{s_i \in \Omega} n(s_i,t)
\end{equation}
which is the key quantity which has to be analysed here.         

 Any microscopic host behavior compatible with the fluid scenario must 
ensure a controlled injection of packets in order to not flooding the network 
(and thus entering the congested state). Note that in a fluid traffic regime 
it is unlikely that packets will be alive for an arbitrarily long time. A
fluidity requirement will necessarily impose a hard constraint on the maximum
time spent by a packet travelling across the network. The simplest way of 
ensuring fluid packet flow is to stop the sources 
emitting new packets when detecting local congestion, that is, when there 
is no empty space in the source neighborhood $C_i$ to fill with new packets. 
If no self-regulation is present, it has been shown that 
there is a sudden and sharp jamming transition from the free to the congested 
state depending on the mean (system) packet generation rate 
$\left\langle \lambda\right\rangle$. At the critical point between the two phases 
the system displays optimum (global) performance.  If self-regulation is allowed, 
it can be shown 
\cite{ValverdeSole2002} that simple traffic source rules are able to self-organize 
the traffic network around a definite mean rate $\left\langle \lambda\right\rangle$,
which scales as the quotient between the average system load $<N>$ and 
the averaged packet latency $\left\langle T \right\rangle$:

\begin{equation}
\lambda _C  \approx \frac{{ < N > }}{{ < T > }}
\end{equation}
where the latency is defined as the time comprised from the 
creation of a packet until its delivering at the destination host. The 
mean latency $<T>$ is averaged over all successfully released 
packets. Following \cite{ValverdeSole2002} we will indicate by $\xi$
the number of local congested neighbors:
\begin{equation}
\xi  = \sum\limits_{k \in C(i)} {\theta \left[ {n(i,t)} \right]} 
\end{equation}
where $\theta[x]=1$ for $x>0$ and zero otherwise. The packet injection 
rate for host i is updated as follows:
\begin{equation}
\lambda _i (t + 1) = \left\{ 
{\begin{array}{*{20}c}
   {\min \left\{ {1,\lambda _i (t) + \mu } \right\}} & {\xi  = 0}  \\
   {\lambda _i (t)} & {0 < \xi  < k_i }  \\
   0 & {\xi  = k_i }  \\
\end{array}} \right.
\end{equation}

where $\mu$ is the so-called driving parameter. The second rule 
allows a particular host to stabilize around a given rate. Traffic 
rate increases conservatively and drops down to zero when all 
neighboring nodes are congested. The reader must be aware 
that the above rules are not intended to be a detailed model of 
real traffic sources. Traffic sources can not be described
with an universal distribution probability. Moreover, it seems 
that a rich variety of distributions (exponential, bi-modal or log-normal)
apply. As we will see later in the paper, it is unlikely that the 
variability of traffic sources will be the cause of the scaling detected 
in Internet traffic. Moreover, this model does not introduce any explicit 
correlations between sources (i.e.: like in active conversation 
between two hosts) so the dynamics can not be the by-product
of pre-defined source correlations. Because the system is poised 
to criticality (the so-called fluid packet flow regime) the global dynamics 
emerge from the collective behavior of its components.  
Ultimate source details are irrelevant in this context. Moreover, although TCP 
is more realistic than our simplified local rules (also much more complex) 
our conjecture is that the large-scale dynamics will not be greatly changed by the local behavior 
of hosts within this fluid flow regime. This idea has been partly answered in a 
recent work \cite{Hohn} in which it has been shown that  IP level traffic (packet dynamics) does 
not depend to any significant extent on the TCP arrival process (host behavior), supporting our view that 
 traffic dynamics is not a consequence of detailed source (host) behavior.

For a homogeneous network with average degree $<k>$ (and finite $<k^2>$) it can be shown, following 
\cite {ValverdeSole2002} that the time evolution of packet density $\Gamma(t)=N(t)/M$ in the limit of 
$H \rightarrow \infty$ is defined by the mean field equation:
\begin{equation}
  {d\Gamma \over dt} = \rho \lambda - {<k> \over D} \Gamma \left (1 - \Gamma \right)
\end{equation}
where $D$ indicates the network diameter (also the average transient time, assuming packets jump 
from node to node with the same average time scale). It is not difficult to show that 
the equilibrium points of the previous equation are given by 
$\Gamma_{\pm} = [1 \pm (1 - 4 \rho \lambda D/ <k>)^{1/2} ]/2$. 
For $\lambda>\lambda_c \equiv <k> / 4D \rho$, 
the fixed points vanish and no finite density exists. In
this ``congested phase'' the density of packets grows without bounds. 
For $\lambda<\lambda_c$, a finite stable density 
\begin{equation}
\Gamma_- =  {1 \over 2} \left [1 - \left ( 
1 - {4 \rho \lambda D \over <k> } \right )^{1/2} \right ] 
\end{equation}
is observable (the other fixed point $\Gamma_+$ is unstable). For the particular
case $\rho=1$ analysed by Fuk\'{s} and Lawnizak \cite{Fuks}, we recover
their critical point $\lambda_c=2/L$ for dynamics taking place on a square lattice, 
i. e. $D=L/2$. 
If a feedback exists between $\lambda$ and $\gamma$, some finite 
equilibrium density $\Gamma^*$ will be achieved in the previous model and thus no divergence will be
allowed to occur. In this case, the mean field model indicates that a scaling 
relation will be observed, i. e. 
\begin{equation}
\lambda \sim \rho^{-1}
\end{equation} 
between the (self-organized) packet release rate and the density of
hosts (which is here the only relevant external parameter). Such a scaling relation was shown to 
occur in the previous model.

The previous mean field calculation was done by assuming that 
a fixed $\lambda$ is beign used. The previous rules actually introduce 
self-regulation of injection rates by traffic. In other words, if 
traffic level $N$ defines an order parameter, it will interact with a control 
parameter ($\lambda$), reducing it when $N$ is large and incresing it when 
low. The feedback between these two key quantities results in a 
state dominated by fluid, but fluctuating traffic with many characteristics 
in common with observed Internet's dynamical patterns. 

In order to explictly consider the feedback between order and control parameters, 
we can consider a new mean field approximation based on the previous local rules. It can be 
shown, asuming finite $H$, that the new set of equations is now:
\begin{equation}
{d\Gamma \over dt} = \rho \lambda \left (1 - {\gamma \over H} \right ) - {<k> \over D} \Gamma 
\end{equation} 
\begin{equation}
{d\lambda \over dt} = \mu (1-\lambda) - {\Gamma \over <k>} 
\end{equation} 
for low density levels (i. e. $\Gamma \ll H$, consistently with a fluid traffic) 
the single fixed point (obtained from $d\Gamma/dt=d\lambda/dt=0$) is 
\begin{equation}
(\Gamma^*, \lambda^*)= \left( {1 \over {<k> \over \rho D} + {1 \over \mu <k>}}, 1-{\Gamma^* \over <k>M} \right )  
\end{equation} 
the Jacobi matrix ${\bf L}$ for the previous set of equations is given by
\begin{equation} 
{\bf L} = \left ( 
\matrix{  {-<k> \over D} &  \rho \cr 
          {-1 \over <k>} &  -\mu  \cr  } \right )   
\end{equation}  
The associated eigenvalues are 
$$
\Lambda_{\pm} = {1 \over 2} \left [ 
- \left ({<k> \over D} + \mu \right ) \pm 
\sqrt{ \left ( {<k> \over D} + \mu \right )^2 - {4 \rho \over <k>} } \right ]
$$
both of them real and negative: the point attractor is globally stable. Numerical simulations of 
the model on a Poissonian graph (where the previous approximation would held) agree with these 
predicted values. Topological features are included only as averaged quantities, here mean degree $<k>$ and 
diameter $D$. However, our interest is to explore the traffic dynamics on a realistic 
network architecture, in order to provide the closest modeling approach under the previous rules. 
It has been shown that the Internet does not display the homogeneous architecture assumed by the 
Poissonian graphs \cite{FFF99}. In the next section the network's topology used in our analysis is presented.
Together with an explicit definition of the routing algorithm, they will complete our model's description.

\begin{figure}
\begin{center}
\includegraphics[scale=1]{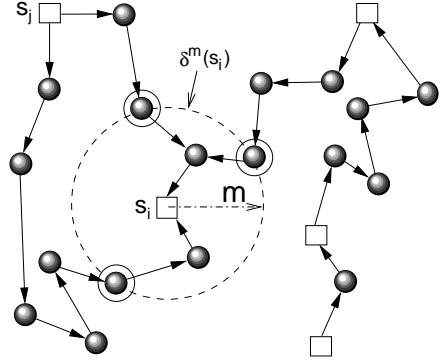}
\vspace{0.2 cm}
\caption{Network dynamics and routing: it involves a given depth of routing $m$ (a path horizon). 
A packet traveling from $s_j$ to $s_i$ within the $\delta^m(s_i)$ domain (of depth $m$) 
is deterministically routed along the shortest path ($d(j,i)\le m$). The packet traverses 
hosts (squares) and routers (circles) indistinctly. Packets traveling outside the $m$-domain (i. e. for $d(j,i) > m$) have 
more than one path choice and perform random walks. As soon as the packet enters into the $m$-domain, the packet 
is deterministically routed along the shortest path. Here we would have $m=2$.} 
\label{fig1}
\end{center}
\end{figure}

\section{Network topology}

In previous analyses \cite{SoleValverde2001} \cite{ValverdeSole2002}
the topology of the graph was chosen to be a lattice. Although this might 
seem a limited topological arrangement, it has been successfully tested 
on real hardware and the presence of a phase transition fully confirmed 
\cite{Bolding}.  Moreover, mesh connected topologies may be the 
most efficient solution at the limit of very large parallel computer 
architectures \cite{Vitanyi}\cite{Bilardi}.

Regular lattices fail to describe Internet topology. 
Remarkably, Internet displays a scale-free architecture \cite{FFF99}. 
The origins and causes of this topology have been the subject of most 
discussion and controversy. It has been show that
most existing Internet generators fall in a very different region of
the phase space where real Internet is located \cite {Yook}. This phase 
space is defined by $(D_f, \sigma, \alpha)$, where every 
parameter defines a major force shaping a different aspect of 
the large-scale Internet topology.  Note that this model does not define 
all detailed correlations observed in Internet and/or the precise 
functional form of Internet's path length and degree distribution
(i.e.: the exact exponent of the scale-free distribution). This 
is a  minimal  set of universal parameters that any realistic Internet 
model  must satisfy and it is a very good approximation to the 
large-scale topology.

        The spatial distribution of Internet nodes is not random. It has
been noticed a strong correlation between fractal distribution of cities
and Internet nodes. The measured fractal dimension is $D_f = 1.5$. When
generating the Internet-like network, the position of nodes is obtained
by sampling a Rayleigh-L\'{e}vy dust of the same fractal dimension. The
likelihood of placing a link between two nodes $s_i$ and $s_j$ depends 
both on the (euclidean) link length $w_{ij}$ and linear preferential attachment:
\[
\Pi (k_j ,d_{ij} ) \approx \frac{{k_j^\alpha  }}{{w_{ij}^\sigma  }}
\]
Note that longer links cost more and thus will be selected with less
probability. Traditional topology generators based on Waxman
model \cite{Waxman} wrongly assume exponential decay instead of 
linear cost decay. On the other side, there is empirical evidence that highly connected
nodes will be linked with higher probability. Increasing $\alpha$ will favor 
linking to nodes with higher degree, while a higher $\sigma$ will penalize 
longer links.  From real measurements \cite{Yook}, it has been identified 
the position of Internet at $D_f=1.5$,  $\sigma=1$ and $\alpha=1$.  Here,
all numerical simulations have been performed using this topological arrangement (figure 1).

\section{Path horizon and routing tables}

An important ingredient when modeling network dynamics is to 
describe the paths followed by packets towards their destinations, 
that is, the routing policy. By properly defining this routing algorithm, 
we will complete our model's definition.

Real routing protocols do not drive packets at random. Instead, they 
try to route packets along the most efficient routes 
(i.e: minimize distance or latency). At the same time, it is unrealistic 
to assume that all packets follow optimal paths because the large amount 
of global information replicated at every single node. 
Clearly, the paths traced by packets in real networks are properly 
characterized as a trade-off between random diffusion and optimal routing. 
This is reflected in the real Internet by its two-level routing organization. 
Nodes are grouped in so-called Autonomous Systems (AS)
and different routing rules apply at each level. Intra-AS routing is based
on shortest path routing but inter-AS routing does not cleary follow
any minimization criteria.

\begin{figure}
\begin{center}
\includegraphics[scale=0.55]{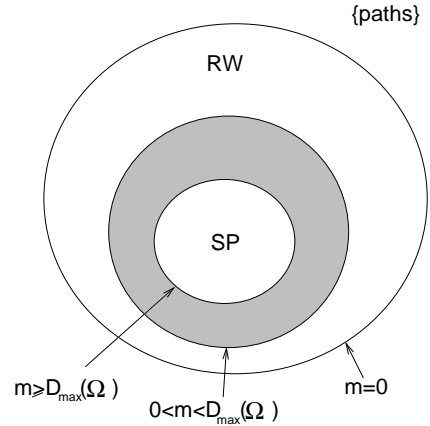}
\vspace{0.2 cm}
\caption{ Hierarchy of path sets defined by the routing policy. Every routing scheme 
explores a fraction of network paths (a subset of the entire path set $P(\Omega)$). Pure 
random walk strategy (m=0) visits all available paths, even shortest paths. 
In fact, random walk travel more frequently along those shortests paths. 
The more restrictive set is associated to pure shortest path routing 
($m \ge [\left\langle d \right\rangle]$), which chooses a small fraction 
of all possible routes on the network.}
\end{center}
\end{figure}

A simple way to explore the cost/efficiency trade-off is by introducing a 
parameter defining the visibility scope of the node (depth of 
routing parameter $m$ or node domain diameter), that is, a 
sphere $\Gamma^{(m)}(s_i)$ of radius $m$ centered at every node $s_i \in \Omega$ (figure 1).  
We allow every node to known every other node at a distance of $m$ hops or less but 
no more. No information will be stored about nodes outside 
the node domain. This idea is indicated in figure 2, where a given 
target node $s_i$ is shown at the center of its $m$-sphere. When a 
foreign node $s_j \in \Omega-\Gamma^{(m)}(s_i)$ send a packet towards $s_i$, 
while moving in the outside of the sphere it performs as a random walk. 
Once the packet hits the boundary of the sphere, $\partial \Gamma^{(m)}(s_i)$, 
it is routed along the shortest path \cite{NOTE1}.

\section{Efficiency and network's exploration}

        The depth of routing parameter m induces a hierarchy of path 
subsets over the entire set of available network paths (see figure 3). 
The random $m=0$ and deterministic $m=M$ routing represent the most
general and restrictive subsets, respectively. Increasing m will progressively reduce the 
randomness at routing decisions. Here, we find a nested collection of 
subsets corresponding to the intermediate situations $0 < m < M$.
The relative size of each subset can be approximated by measuring the 
fraction of nodes F visited by packets:
\begin{equation}
F = \frac{1}{M}\sum_{i = 1}^M \theta \left[ {A_i } \right]
\end{equation}
where $A_i$ is the probability of visiting the node $s_i$. This term depends
both on the network dynamics and the geometrical situation of the node 
within the network. Considerable effort has been devoted to characterize 
the likelihood of visiting a node in a purely static fashion. In this context, 
two relevant centrality (or 'load' \cite{Noh}\cite{Goh2003}) measures are 
Random Walk Centrality \cite{Newman2003} and Betweenness Centrality
\cite{Freeman77} which certainly represent the two extremes of our
hierarchy of path subsets. Anyway, the correlation of these measurements
with dynamic centrality $A_i$ is weak, to say the least.

In figure 4 the numerically measured $F(m)$ for a finite range of 
$m$  values is shown. This fraction equals 1 for 
random routing $0 \le m < [\left\langle d \right\rangle]$, 
where $\left\langle d \right\rangle$ is the averaged shortest path length 
and $[x]$ denotes the integer part of x. Random routing does 
not discard any network node. A sharp transition takes place from this point 
$ [\left\langle d \right\rangle] < m \le M$ and a large fraction of network nodes 
(about 45 percent) are never visited by deterministic routing. This severly 
restricts the diversity of routes and yields a load-insensitive system 
\cite {Lorenz}. The system 
requires a certain degree of noise in order to avoid sending packets 
through already collapsed nodes while it is enabled to choose less 
optimal but free routes.  Note that depth of routing parameter $m$ 
defines an order parameter because quantifies the degree of order 
existing in the system. The existence of an order parameter confirms 
the critical nature of Internet traffic.

\begin{figure}
\begin{center}
\includegraphics[scale=0.7]{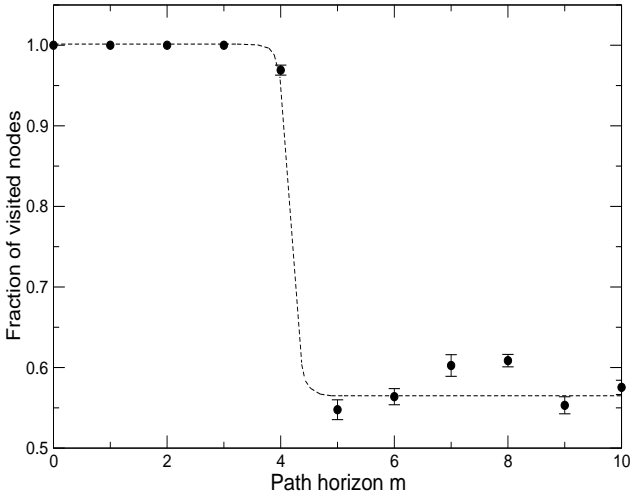}
\vspace{0.2 cm}
\caption{The fraction of visited nodes depends on the amount of
order defined by the depth of routing parameter. Determinism sharply 
constraints the routing paths. Every point was obtained averaging 
over an ensemble of four different networks and every single network 
was simulated in five different host configurations. Network 
parameters: M = 512, $<k>=4$, $\rho = 0.1$, $\mu = 0.01$, H = 10, 
$D_f= 1.5$, $\sigma = 1$, $\alpha = 1$ and $T = 7 \times 10^5$ steps.} 
\label{fig4}
\end{center}
\end{figure}

Numerical simulations have shown that flow is maximized at 
the order-disorder transition point $m= [\left\langle d \right\rangle]$.
This can be observed in network throughput reaching a maximum 
at this point (see figure 5). Throughput is defined by the quotient of 
the number of successfully released packets and the sum of all 
packets generated during the simulation. In the next section,
we will show that several real statistics can be reproduced by the 
system dynamics poised to criticality.

\begin{figure}
\begin{center}
\includegraphics[scale=0.55]{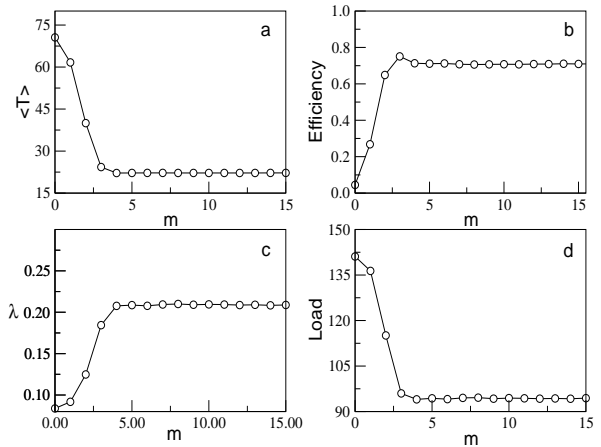}
\vspace{0.2 cm}
\caption{Exploring the network traffic dependency on depth of routing parameter ($m$). The lines connect 
the points for illustrating purposes only. (A) Mean latency is considerably reduced when enough 
system information is given to individual nodes. Most packets are forwarded along the minimum number of hops. 
(B) Global throughput is optimal at the critical point when path horizon is about the network 
diameter.  (C) Mean packet rate also drops down at intermediate value. (D) Mean workload also
experiences sudden transition from heavy to light load. Network parameters: M=250, $<k>=4$,  $D_f = 1.5$,  $\sigma = 1$, $\alpha = 1$. 
Simulation parameters: $\rho = 0.1$,  $\mu = 0.01$,  H = 10,  $T=10^5$ steps. The shape of these plots 
does not depend on network size, that is, the optimal point is always found at $m = [\left\langle d \right\rangle]$.} 
\label{fig5}
\end{center}
\end{figure}

\section{Average stretch}

        The average stretch $s$ measures the efficiency of routing by comparing 
the number of hops $h$ traversed by a packet to the shortest path distance $d$ 
between source and destination:

\[
s = \frac {h} {d}
\]

Compact routing schemes minimize the average stretch while maintaining
the size of routing tables small \cite{Cowen}. Reducing the average 
strech will progressively raise up the memory requirements at each router.
Assuming Thorup-Zwick (TZ) compact routing scheme \cite{TZ2001}, the average stretch can 
be expressed as a function of the distance distribution and the graph size $M$ only: 
$\left \langle s \right \rangle = f(\left\langle d \right\rangle, \sigma_d)$ \cite{Krioukov}.  
This TZ scheme ensures a nearly optimal 
lower memory upper bound for $\left\langle s \right\rangle=3$ in generic networks.
For scale-free networks, TZ achieves lower bounds. In particular, for 
the Internet interdomain graph (Autonomous Systems) with degree distribution 
exponent $\gamma \approx 2.1$ and $M = 10^4$, the TZ average 
stretch $\left\langle s \right\rangle \approx 1.14$\cite{Krioukov}. Also, 
the average number  of entries in the routing tables is approximately 52. It 
turns out that $\left\langle s \right\rangle(\left\langle d \right\rangle, 
\sigma_d)$ surface has unique minimums. 
Strikingly,  the points corresponding to Internet distance distribution are 
very close to them \cite{Krioukov}. This suggests that Internet topology 
is shaped by some hidden optimization criteria. Anyway, TZ scheme is not 
a realistic Internet interdomain routing scheme because assumes that global 
topology view is available. 
        We have numerically measured the average stretch and the average fraction of neighbours
at distance $m$ for our routing scheme (see figure 6). Note that the critical path horizon  
$m = [\left\langle d \right\rangle]$ is very close to the minimum average stretch.  

\begin{figure}
\begin{center}
\includegraphics[scale=0.7]{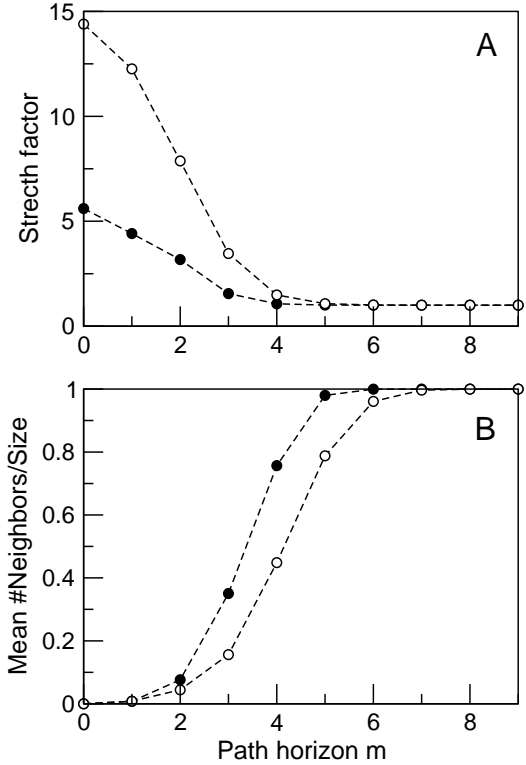}
\vspace{0.1 cm}
\caption{(A) Average stretch (B) Average fraction of nodes inside $m$-domain is
an estimation of the amount of information required by each node. Here open circles correspond 
to the Internet's nodel. For comparison,
the simulation has been repeated in a poissonian network of same size and 
mean degree (filled circles). Network parameters: $M=500$, $<k>=4$,  $D_f = 1.5$,  
$\sigma = 1$, $\alpha = 1$. Simulation parameters:  
$\rho = 0.1$, $\mu = 0.01,  H = 10,  T = 5 \times 10^4$ timesteps. 
Measurement window = 200. The distributions were obtained from an statistical ensemble 
of three networks, every network simulated three times with hosts located at different configurations.} 
\label{fig10}
\end{center}
\end{figure}

\section{Network fluctuations and performance}

        In order to test the goodness of the model presented here, it is interesting
to compare some real Internet statistics with their respective model measurements.
In particular, those measurements will be collected for the model at the critical path horizon,
that is, when $m = [\left\langle d \right\rangle]$. 

        When understanding the competition between a network's internal 
collective dynamics (i.e: Internet traffic) and external environmental changes 
(i.e: traffic sources or host behaviour), it is useful to study the relationship 
between the mean flux and the size of fluctuations around the average 
\cite{Fluctuations}. Previous explanations of ocurrence of self-similarity in
traffic networks are based on the superposition of many and high-variable (infinite
variance) sources \cite{Willinger}. This point of view discards the effects of 
the system's collective dynamics and considers that Internet is an externally 
driven  system. Real data shows that Internet dynamics 
can not be simply reduced to the behaviour of traffic sources. 
        
        Let us define the incoming (outgoing) flux $f_i$ as the amount of packets 
received (forwarded) at router $s_i$ during a given and fixed period of time. For
every router, compare the average flux $\left\langle f_i \right\rangle$ with the 
dispersion $\sigma_i$ around the mean. It has been noticed that for several real 
systems the following scaling relation holds:

\[
\sigma  \approx \left\langle f \right\rangle ^\alpha  
\]
        
where $\alpha$ is an exponent which can take the values of 1/2 and 1 \cite{Fluctuations}.
This suggests that real systems can be classified in two main classes depending on the
value of this exponent. The relevant exponent in this context is $\alpha = 1/2$, which
has been observed in daily traffic measurements at 374 geographically distinct Internet 
routers \cite{Fluctuations}. Systems exhibiting the 1/2 exponent are representative of
endogeneous dynamics, that is, determined by the system's internal collective 
fluctuations.  Moreover, measurements on our model reproduce the same exponent 
(see figure 7).

\begin{figure}
\begin{center}
\includegraphics[scale=0.65]{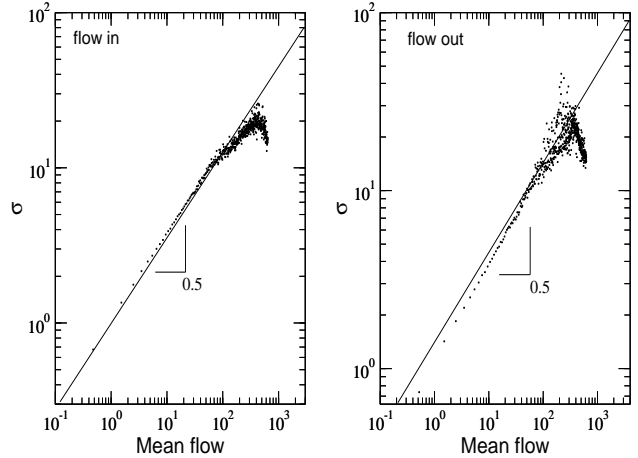}
\vspace{0.2 cm}
\caption{The relationship between fluctuations and the average incoming node flux (a similar distribution holds 
for average outgoing router flux). The plot shows that both quantities are related by a power law of  
exponent 1/2, which is consistent with the measurements from Internet routers. Network parameters: 
M=500, $<k>=4$,  $D_f = 1.5$,  $\sigma = 1$, $\alpha = 1$. Simulation parameters:  $\rho = 0.1$, $\mu = 0.01$,  H = 10,  T = 10000 steps. 
Measurement window = 200. The distributions were obtained from an statistical ensemble of 10 networks, every 
network simulated 5 times with hosts located at 5 different configurations.} 
\label{fig7}
\end{center}
\end{figure}

        A measure of Internet end-to-end performance is the normalized latency time $\tau$ 
\cite{Vespignani} defined as the quotient between latency time $L$ (measured as 
Round-Trip-Time) and geographical distance $w$:

\[
 \tau = \frac{L}{w}
 \]
 
There are several factors governing $\tau$. First, propagation rate is finite. There is a minimum 
delay because packets can not move faster than speed of light.  Second, the number of nodes 
traversed departs from the minimum found along the geodesic path from node to destination. 
And third, packets spent some time at every intermmediate node because queueing delays. 
Define $\tau_{min}=L_{min}/w$ as the normalized latency without taking into acount queue 
delays and $\tau_{av}=L_{av}/w$ as the normalized latency while considering all factors 
affecting packet latency.  

Their probability distributions have been measured from two years of 
PingER \cite{PingER} data and follow power-law scaling with stable exponents of about -3.0 for $\tau_{min}$ and 
-2.5 for $\tau_{av}$ \cite {Vespignani}. Numerical simulations on the model poised to criticality 
reproduced the power-law $P(\tau_{av})$ probability distribution. The exponent of the 
distribution is about -2.45, which is very close to real observations (see figure 8). 
The previous results seems to be quite robust and independent of most model 
parameters changes. An interesting thing to note is that current model cannot 
reproduce the existing correlation between $\tau_{min}$ and $\tau_{av}$ and 
geographical distance $w$ reported by previous studies \cite{Vespignani}. The reason might be 
that propagation rate is (unrealistically) assumed to be infinite in the model.  
It might be that this factor has little influence in shaping the $\tau_{av}$ probability 
distribution. Moreover, it is worth noting that any deviation of the order parameter from the critical regime 
resulted in an exponential distribution for $\tau_{av}$, reinforcing the view of an optimally 
efficient and self-organized Internet.

\begin{figure}
\begin{center}
\includegraphics[scale=0.7]{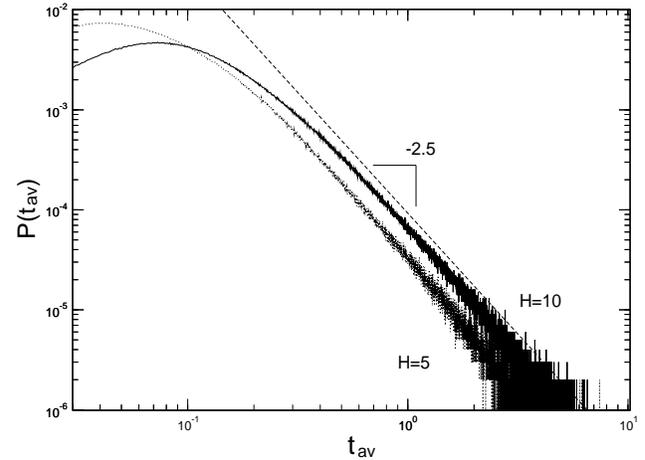}
\vspace{0.2 cm}
\caption{
Normalized latency $\tau_{av}$ distributions at the critical point $m = [\left\langle d \right\rangle]$
follows
power law of exponent $ \sim -2.45$ and fits very well the real measurements (see text). Here, we plot the 
distributions for $H= 5$ (shaded line) and $H=10$ (continuous line) showing that the long tail does not depend on the 
maximum queue size. Any deviation from $m = \left\langle d \right\rangle$ results in an exponential 
distribution, deviating from real observations. Network parameters: $M = 1000$, $<k>=4$, 
$D_f=1.5$,  $\sigma=1$, $\alpha= 1$. Simulation parameters: $\rho=0.1$, $\mu = 0.01$, 
$H=5,10$ and $T=5 \times 10^5$ steps. The distributions were obtained from an statistical ensemble of 
10 different networks and every network was simulated five times with different host
arragements.} 
\label{fig6}
\end{center}
\end{figure}

\section{Discussion}

In this paper we have shown that a simple model of traffic dynamics incorporating 
the appropriate Internet's network topology is able to recover several 
statistical features of real traffic. More important, we have seen that 
the routing algorithms can take advantage of the small-world structure of the 
web by reaching a critical path horizon close to the network's average path length. In doing so, 
a highly efficient system is reached at low cost: routing strategies only need to 
consider a small depth. Once the $m$ parameter reaches the network's diameter, no further 
information is required to properly reach the target. A full-system deterministic routing 
strategy is actually unnecesary and would be too costly. Instead, the constraints imposed 
by network's architecture allow to exploit the implicit information defined by the 
small-world architecture.

        It might be that Internet evolves in a way that throughput or
global performance is maximized. This trend is constrained within the
limits of available communication resources. Inside this regime,
Internet is shaped in order to provide better response. This is reflected
in the optimal and lower average stretch observed in real Internet,
which is close to global minimum \cite{TZ2001}. This suggest hat some
hidden optimization is at work. Clearly, an unresponsive system will be 
no good. How a distributed collection of designers were able to 
define this globally efficient infrastructure is a question that deserves
attention.

\begin{acknowledgement}
This paper is dedicated to the memory of Per Bak. 
This work was supported by a grant BFM2001-2154 and by the Santa Fe Institute.  
\end{acknowledgement}

{}


\begin{thebibliography}{}

\bibitem{Yook}
S-H. Yook, H. Jeong, and A-L. Barab\'{a}si ,
Proc. Natl Acad. Sci. USA, {\bf 99}, 13382, (2002)

\bibitem{Huberman}
B. A. Huberman and R. M. Lukose, Science, {\bf 277}, (1997)

\bibitem{Wiener}
N. Wiener, {\em Cybernetics}, John Wiley and Sons, New York (1949)

\bibitem{Ogielski}
James H. Cowie, David M. Nicol, Andy T. Ogielski,  Computing in Science and Engineering, 
 {\bf 1}(1), 42-50 (1999)
 
\bibitem{WillingerPNAS}
W. Willinger, R. Govindan, S. Jamin, V. Paxson, and S. Shenker,
PNAS {\bf 99} (Suppl. 1), 2573-2580 (2002)
 
\bibitem{SoleValverde2001}
R. V. Sol\'{e} and S. Valverde, Physica A, {\bf 289}, 595 (2001)

\bibitem{ValverdeSole2002}
S. Valverde and R. V. Sol\'{e}, Physica A, {\bf 312}, 636 (2002)

\bibitem{Hohn}
N. Hohn, D. Veitch and P. Abry, 
In; ACM/SIGCOMM Internet Measurement Workshop, Marseille, France. pp. 63-68. (2002)

\bibitem{Fukuda}
K. Fukuda, {\em A Study of Phase Transition Phenomena in Internet Traffic},
PhD Thesis, Keio Univ. (1999)

\bibitem{Fluctuations}
M. Argollo de Menezes and A-L. Barab\'{a}si, cond-mat/0306304 (2003)

\bibitem{Willinger}
W. Willinger, M. S. Taqqu, R. Sherman and D. V. Wilson, IEEE\slash ACM 
Trans. on Networking, {\bf 5}(1), 71-86 (1997)

\bibitem {Fuks}
H. Fuk\'{s} and A. T. Lawniczak, adap-org/9909006 (2001)

\bibitem{FFF99}
M. Faloutsos, P. Faloutsos and C. Faloutsos, ACM SIGCOMM 
{\bf 29}(4), 251-262, (1999)

\bibitem {Bolding}
K. Bolding, M. L. Fulgham and L. Snyder, Tech. Rep CSE-94-02-04 (1994)

\bibitem{Vitanyi}
P. M. B. Vit\'{a}nyi, SIAM J. Comput. 17,{\bf 4}, 659-672, (1988)

\bibitem{Bilardi}
G. Bilardi and F. P. Preparata, CS-93-20, Dept. Comp. Sci. , Brown Univ. (1993)

\bibitem{Waxman}
B. Waxman, IEEE J. Selec. Areas Commun., SAC-{\bf 6}(9), 1617-1622, (1988)

\bibitem {Romu}
A. V\'{a}zquez, R. Pastor-Satorras, and A. Vespignani, cond-mat/0206084 (2002)

\bibitem{NOTE1}
When the link decision is ambiguous (more than one link can be 
selected) the less visited link until the moment is chosen (this 
could be implemented by maintaining a counter of the number 
of packets forwarded through the link).

\bibitem{Noh}
J. Dong Noh and H Rieger, cond-mat/ 0307719 (2003)

\bibitem{Goh2003}
K.-I. Goh, B. Kahng, and D. Kim, {\em Traffic and Granular Flow '01}, Springer, Berlin (2003)

\bibitem{Newman2003}
M. E. J. Newman,  cond-mat/0309045 (2003)

\bibitem{Freeman77}
L. C. Freeman, Sociometry {\bf 40}, 35 (1979)

\bibitem{Lorenz}
D. H. Lorenz, A. Orda, D. Raz and Y. Shavitt, TR-2001-17, DIMACS (2001)

\bibitem{Cowen}
L. J. Cowen, Proc. of the 10th Annual ACM-SIAM Symp. on Discrete Algorithms, (1999)

\bibitem {TZ2001}
M. Thorup and U. Zwick, Proc. 33th Annual ACM Symposium on Theory of 
Computing (SPAA), 1-10 (2001)

\bibitem {Krioukov}
D. Krioukov, K. Fall and X. Yang, cond-mat/0308288  (2003)

\bibitem{Vespignani}
R. Percacci and A. Vespignani, cond-mat/0209619 (2002)

\bibitem{PingER}
Internet End-to-end Performance Monitoring, http://www-iepm.slac.stanford.edu.


\end{thebibliography}
\end{document}